\documentclass{iopconfser}
\usepackage{amsmath}
\usepackage[mathscr]{euscript}
\usepackage{amssymb}
\usepackage {mathtools}
\usepackage{hyperref}
\newcommand{\proptorev}{\mathrel{\reflectbox{$\propto$}}}
\usepackage[T1]{fontenc}
\usepackage{tipa}

\begin{document}

\title{Black Hole Scattering and Integrability: \\A Hyperboloidal Approach}

\author{Corentin Vitel}

\affil{Université Bourgogne Europe, CNRS, IMB UMR 5584, F-21000 Dijon, France}

\email{corentin.vitel@u-bourgogne.fr}

\begin{abstract}
Integrability structures are known to play a key role in one-dimensional scattering. In the Schwarzschild gravitational context, the analysis emphasizing the role of the so-called Darboux covariance and its intimate connection with KdV conserved quantities was recently introduced by Lenzi \& Sopuerta. In a second stage, together with Jaramillo, this led in particular to the identification of the structural role of the “KdV-Virasoro-Schwarzian derivative” triangle in this problem. Such a gravitational scattering description dwells naturally on a Cauchy foliation of the spacetime. In the following, we first review --- for the Schwarzschild background --- this problem in a hyperboloidal foliation scheme, where the infinitesimal time generator of the dynamics is a non-selfadjoint operator. Then, we explore the underlying integrability features through a Lax-pair formulation. Specifically, the main results presented here are i) the explicit proposal of a \textit{weak} Lax-pair, valid under suitable conditions involving fields at null infinity, with ii) the construction of the associated infinite sequence of isospectral flows. From a broader perspective, the very form of the non-selfadjoint infinitesimal time operator, which neatly separates into two components corresponding to bulk and boundary structures, paves the way for the description of the gravitational dynamics in terms of a “semi-direct action” of bulk degrees of freedom onto boundary degrees of freedom. This is akin to the “wave-mean flow” approach for black hole strong-gravity dynamics recently proposed in this line of research.
\end{abstract}

\section{A brief overview of $1+1$ Integrable Structures in a Cauchy slicing}



In the Black Hole Perturbation Theory (\textit{BHPT}) framework, the \textit{Schwarzschild} \textit{master equations} \cite{ref1} are wave-like equations with a potential\footnote{Interpreted as an effective potential encoding the influence of the background spacetime on linear perturbations.}, whose spectral problem is given by a \textit{Schrödinger operator} \cite{ref2}
\begin{equation}
    \big(-\partial_t^2+\partial_x^2-V_l^{\text{even/odd}}\big)\Phi^{\text{even/odd}}=0 \Rightarrow \mathcal{L}_V\phi^{\text{even/odd}}\equiv\big(\partial_x^2-V^\text{even/odd}_l\big)\phi^\text{even/odd}=-\omega^2\phi^\text{even/odd}\hspace{0.1cm} ,  
    \label{1}
\end{equation}
where $x$ refers to the tortoise coordinate while $\Phi^{\text{even/odd}}=e^{i\omega t}\phi^{\text{even/odd}}$ represents the \textit{master function}. 

Traditionally (e.g., \cite{ref3,ref4}), one associates to $\mathcal{L}_V$ a non-linear operator $\mathcal{P}_V$ such that \textit{i)} their commutator yields a non-linear (dispersive) \textit{integrable} PDE and \textit{ii)} the \textit{isospectrality} of the spectrum is preserved along the flow. We say $(\mathcal{L}_V,\mathcal{P}_V)$ defines a \textit{Lax pair}. 
Specifically, one deforms Eq.\eqref{1} with respect to some $\sigma$-time (distinct from the time $t$) so that $\phi_{,\sigma}=\mathcal{P}_V\phi$. The \textit{Lax equation} $\partial_\sigma\mathcal{L}_V=[\mathcal{P}_V,\mathcal{L}_V]$ then expresses the compatibility between the spectral problem Eq.\eqref{1} and the $\sigma$-deformation of the solution $\phi$\footnote{\textit{Isospectrality} then means $\omega_{,\sigma}=0$ when the Lax equation holds.}. 

The case $\mathcal{P}_V=-4\partial_x^3+6V\partial_x+3V_{,x}$ leads to the well-known \textit{Korteweg-de Vries} (\textit{KdV}) equation 
\begin{equation}
    \partial_\sigma\mathcal{L}_{V}=[\mathcal{P}_V,\mathcal{L}_V] \Leftrightarrow V_{,\sigma}-6VV_{,x}+V_{,xxx}=0 \hspace{0.1cm}.
    \label{2}
\end{equation}
KdV exemplifies integrability: from the \textit{r-matrix} formalism (e.g., \cite{ref5,ref6}) and the \textit{Virasoro algebra} (e.g., \cite{ref7,ref8,ref9}) to \textit{pseudo-differential operators} and related \textit{Hamiltonian structures} (e.g., \cite{ref10}).

\section{Lax pair in Hyperboloidal Approach through the \textit{weak} adjoint viewpoint}

Using a Cauchy description of spacetime, the master functions blow up asymptotically \cite{ref11}, hindering the identification of the algebraic structures at null infinity $\mathscr{I}^+$ \cite{ref12}. In this work, we adopt a \textit{hyperboloidal} foliation to smoothly connect the event horizon $\mathcal{H}$ with $\mathscr{I}^+$ via spacelike hypersurfaces \cite{ref13}. 
\\
More concretely, the coordinates $(\tau,\xi)$ implement the \textit{compactified hyperboloidal slicing} through $t(\tau,\xi)=\tau-h(\xi)$ and $x(\tau,\xi)=g(\xi)$. The hyperboloidal time function $h$ ensures that the master functions remain regular at these new boundaries while $g$ compactifies the real line onto the domain $[a,b]$.

By introducing a first-order reduction ($\psi:=\partial_\tau\phi$), the Schrödinger dynamics writes in matrix form 
\begin{equation}
    \partial_\tau\mathbf{u}=i\mathbb{L}\mathbf{u}, \hspace{0.1cm} \mathbb{L}=\frac{1}{i}\begin{pmatrix}
        0 & 1 \\ \mathcal{L}_1 & \mathcal{L}_2
    \end{pmatrix} \hspace{0.1cm} \text{with} \hspace{0.1cm}
    \begin{cases}
        \mathcal{L}_1=\frac{1}{p(\xi)w(\xi)}\big[D_\xi^2-\tilde{V}\big] \\
        \mathcal{L}_2=\frac{1}{p(\xi)w(\xi)}\big[2\gamma(\xi)D_\xi+\gamma_{,\xi}(\xi) \big]
    \end{cases} \hspace{0.1cm} ,
    \label{3}
\end{equation}
where $\mathbf{u}^\top=(\phi,\psi)^\top$ and $p$, $w$, $\gamma$ are functions derived from $\partial_\xi g$ and $\partial_\xi h$ \cite{ref12}. We set $D_\xi:=p(\xi)\partial_\xi\equiv _{,\xi}$.
The spectral problem $\mathbb{L}\mathbf{u}=\omega\mathbf{u}$ leads to a quadratic spectral problem in $\omega$, with $\gamma$-dependence
\begin{equation}
    \hat{\mathbb{L}}(\omega)\phi\equiv\phi_{,\xi\xi}+2\omega\gamma\phi_{,\xi}-\big[(1-\gamma^2)\omega^2-\omega\gamma_{,\xi}+\tilde{V} \big]\phi=0\hspace{0.1cm} .
    \label{4}
\end{equation}
We seek the underlying \textit{integrable structure(s)} of the above system through a Lax-pair formulation. 

Due to energy loss at the boundaries \cite{ref12,ref14}, the time-generator $\mathbb{L}$ is not \textit{self-adjoint}. In fact, with an appropriate choice of a so-called \textit{energy scalar product} \cite{ref14}, the adjoint operator takes the form 
\begin{equation}
    \mathbb{L}^\dagger=\frac{1}{i}\begin{pmatrix}
        0 & 1 \\ \mathcal{L}_1 & \mathcal{L}_2+\mathcal{L}_2^\partial
    \end{pmatrix} \hspace{0.1cm} \text{with} \hspace{0.1cm} \mathcal{L}_2^\partial=2\frac{\gamma}{w}\bigg(\delta(\xi-a)-\delta(\xi-b)\bigg)\hspace{0.1cm} .
    \label{5}
\end{equation}
We may therefore also consider the \textit{adjoint spectral problem} $\mathbb{L}^\dagger\underline{\mathbf{u}}=\bar{\omega}\underline{\mathbf{u}}$, $\underline{\mathbf{u}}^\top=(\underline{\phi},\underline{\psi})^\top$, and set $\lambda:=\bar{\omega}$. \textit{A priori} there are two Lax-pair problems: either find some $\mathbb{P}$ such that $\partial_\sigma\mathbb{L}=[\mathbb{P},\mathbb{L}]$, or find some $_{\text{ad}}\mathbb{P}$ such that $\partial_\sigma\mathbb{L}^\dagger=[_{\text{ad}}\mathbb{P},\mathbb{L}^\dagger]$ \footnote{Even if a Lax operator $\mathbb{P}$ exists, its adjoint $\mathbb{P}^\dagger$ does not necessarily defined a Lax operator for $\mathbb{L}^\dagger$.}. In what follows, we focus on the latter --- specifically, on a \textit{weak} version of it. 

\textbf{Proposition} \textit{The weak adjoint spectral problem --- defined as the integration of $\mathbb{L}^\dagger\underline{\mathbf{u}}=\lambda\underline{\mathbf{u}}$ with respect to the measure $wd\xi$ --- leads to an energy dependent Schrödinger operator}
\begin{equation}
    \underline{\phi}_{,\xi\xi}+\big[\lambda^2(\gamma^2-1)-\lambda\gamma_{,\xi}-\tilde{V}\big]\underline{\phi}=0 \hspace{0.1cm} 
    \Leftrightarrow \hspace{0.1cm}^{w}\hat{\mathbb{L}}^\dagger\underline{\phi}\equiv\big(\alpha D_\xi^2+u\big)\underline{\phi}=0\hspace{0.1cm} ,
    \label{6}
\end{equation}
\textit{with $\alpha=\sum_0^2\alpha_k\lambda^k$, $u=\sum_0^2u_k\lambda^k$, where $\alpha_0=1$, $\alpha_1=\alpha_2=0$ and $u_0=-\tilde{V}$, $u_1=-\gamma_{,\xi}$, $u_2=\gamma^2-1$.}

Motivated by Antonowicz \& Fordy \cite{ref15,ref16,ref17,ref18}, we study $\sigma$-time flows induced by the deformation 
\begin{equation}
    \underline{\phi}_{,\sigma}={}^w_\text{ad}\hat{\mathbb{P}}\underline{\phi}\equiv\bigg(\frac{1}{2}PD_\xi+Q\bigg)\underline{\phi}\hspace{0.1cm},
    \label{7}
\end{equation}
where $P$,$Q$ are functions of the $u_k$, their derivatives and the spectral parameter $\lambda$. The compatibility condition ensuring integrability of the system defined by Eq.$\eqref{6}$ and Eq.$\eqref{7}$ is given by the Lax equation 
\begin{equation}
    {}^w\hat{\mathbb{L}}^\dagger_{,\sigma}=\big[{}^w_\text{ad}\hat{\mathbb{P}},{}^w\hat{\mathbb{L}}^\dagger \big]\hspace{0.1cm} .
    \label{8}
\end{equation}
Substituting the elements above, we fix $Q$ such that $P_{,\xi\xi}+4Q_{,\xi}=0$, which yields 
\begin{equation}
    u_{,\sigma}=\bigg(\frac{1}{4}\alpha D_\xi^3+\frac{1}{2}(uD_\xi+D_\xi u)\bigg)P\equiv J_uP=\bigg(\sum_{k=0}^2J_k\lambda^k\bigg)P\hspace{0.1cm},
    \label{9}
\end{equation}
with $J_k=1/4\alpha_kD_\xi^3+1/2(u_kD_\xi+D_\xi u_k)$. For convenience, let us expand $P$ as $P=\sum_{k=0}^mP_{m-k}\lambda^k$ for arbitrary $m\geq1$, with $P_k=0$ for $k<0$. Then, Eq.\eqref{9} decomposes into two sets of equations 
\begin{equation}
    J_0P_{k-2}+J_1P_{k-1}+J_2P_k=0, \hspace{0.1cm} k=0,1,\cdots,m-1 \hspace{0.1cm},
    \label{10}
\end{equation}
\begin{equation}
    \begin{cases}
        (\gamma^2)_{,\sigma_m}\proptorev u_{2,\sigma_m}=J_0P_{m-2}+J_1P_{m-1}+J_2P_m \\
        (\gamma_{,\xi})_{,\sigma_m}\proptorev u_{1,\sigma_m}=J_0P_{m-1}+J_1P_m \\
        \tilde{V}_{,\sigma_m}\proptorev u_{0,\sigma_m}=J_0P_m
    \end{cases} \hspace{0.1cm} ,
    \label{11}
\end{equation}
where $P_k$, $k\in\{0,\cdots,m-1\}$, are determined recursively starting from $P_0\in\text{Ker}J_2$. The last three equations define the \textit{equations of motion} (\textit{e.o.m.}). Thus, we obtain an \textit{infinite} sequence of \textit{isospectral flows} whose last coefficient $P_m$ remains unconstrained for a given $m$. Isospectrality holds by construction. 

One checks $\gamma\equiv 0$ (i.e. $\mathbb{L}^\dagger=\mathbb{L}$, $u_2=-1$) recovers the \textit{hyperboloidal KdV hierarchy}.
In general, since $u_{0,\sigma_m}=\tilde{V}_{,\sigma_m}$, we want to \textit{ensure that the bulk integrable dynamics is properly captured}. Thus, we fix
\begin{equation}
    P_m=\frac{\delta h_m}{\delta u_0}\hspace{0.1cm}, \hspace{0.1cm} \text{where each $h_m$ is a bulk KdV conserved quantity,}
    \label{12}
\end{equation}
namely: $h_1=\frac{1}{2}u_0^2$, $h_2=u_0^3+\frac{1}{2}u_{0,\xi}^2\cdots$ As an example, for $m=1$, we must have $P_1=u_0=\tilde{V}$, and $J_0\tilde{V}=\tilde{V}_{,\xi\xi\xi}-6\tilde{V}\tilde{V}_{,\xi}$ coincides with the right hand-side of the hyperboloidal KdV equation. 

The e.o.m. reveal \textit{3 distinct degrees of freedom} (\textit{d.o.f.}): we emphasize that $\gamma$ should be distinguished from $\gamma_{,\xi}$.

Furthermore,
$\mathcal{U}_{{}^{w}\hat{\mathbb{L}}^\dagger}=\begin{pmatrix}
    0 & 1 \\ -u & 0
\end{pmatrix}$ and $\mathcal{V}_{{}^{w}_\text{ad}\hat{\mathbb{P}}}=\begin{pmatrix}
    -1/4P_{,\xi} & 1/2P \\ -1/2(-1/2P_{,\xi\xi}+Pu) & 1/4P_{,\xi}
\end{pmatrix}$ give a \textit{zero-curvature representation} 
\begin{equation}
    \partial_\sigma\mathcal{U}_{{}^{w}\hat{\mathbb{L}}^\dagger}-D_\xi\mathcal{V}_{{}^{w}_\text{ad}\hat{\mathbb{P}}}+\text{ad}_{\mathcal{U}_{{}^{w}\hat{\mathbb{L}}^\dagger}}\mathcal{V}_{{}^{w}_\text{ad}\hat{\mathbb{P}}}=0\hspace{0.1cm}.
    \label{13}
\end{equation}
Hence, we may expect the existence of conserved quantities attached to such a weak adjoint problem. 

\section{A preliminary study of the Hamiltonian structure(s)}

The previous e.o.m. may be written with respect to a $(2+1)\times(2+1)$ matrix differential operator $\tilde{\mathbf{B}}$
\begin{equation}
    \mathbf{u}_{,\sigma_m}:=\begin{pmatrix}
        u_0 \\ u_1\\ u_2
    \end{pmatrix}_{,\sigma_m}=\tilde{\mathbf{B}}\mathbf{P}^{(m)}, \hspace{0.1cm} \text{where} \hspace{0.1cm} \tilde{\mathbf{B}}=\begin{pmatrix}
        0 & 0 & J_0\\ 0 & J_0 & J_1 \\ J_0 & J_1 & J_2
    \end{pmatrix}\hspace{0.1cm} \text{and} \hspace{0.1cm} \mathbf{P}^{(m)}=\begin{pmatrix}P_{m-2}\\P_{m-1}\\P_m=\delta h_m/\delta u_0\end{pmatrix}  \hspace{0.1cm} .  
    \label{15}
\end{equation}

One shows that such $\tilde{\mathbf{B}}$ is a well-defined \textit{Hamiltonian operator}\footnote{Skew-adjointness is straightforward while Jacobi identity requires a few cumbersome but trivial computations.} which provides us with a first \textit{Hamiltonian structure} $\{\cdot,\cdot\}_{\tilde{\mathbf{B}}}$. For $m=1$, $\mathcal{H}_{m=1}:=h_1+P_0^{-1}$ is a \textit{Hamiltonian functional} in the sense that $\mathbf{u}_{,\sigma_1}=\tilde{\mathbf{B}}\delta\mathcal{H}_{m=1}$. Similarly for $m=2$, $\mathcal{H}_{m=2}:=h_2-2P_1P_0^{-2}$ leads to $\mathbf{u}_{,\sigma_2}=\tilde{\mathbf{B}}\delta\mathcal{H}_{m=2}$. One can then ask: Does it hold $\forall m\geq1, \mathbf{u}_{,\sigma_m}=\tilde{\mathbf{B}}\delta\mathcal{H}_m$? And does there exist extra Hamiltonian structure(s)?\footnote{ If so, we could answer positively the first question using well-known results on Hamiltonian structures in infinite dimensional settings.}

\section{Conclusions}

In the context of BHPT for spherically symmetric BH in Schwarzschild spacetime, we have used hyperboloidal foliations to circumvent the divergences of the master functions at the bifurcation sphere and at spatial infinity by taking into account the spacetime asymptotics. 
This has led to a clean separation between bulk and boundary contributions in the resulting time generator of the dynamics. 

We introduced the \textit{weak} adjoint spectral problem, for which both Lax and zero-curvature pairs exist, enabling us to exhibit an infinite sequence of isospectral flows characterized by 3 d.o.f. This was made possible by fixing the last term in the polynomial expansion of the Lax pair in order to incorporate the hyperboloidal bulk KdV structures. 

From a broader perspective, our aim is to understand such dynamics in $\tilde{V}$, $\gamma$ and $\gamma_{,\xi}$ --- identified as \textit{slow} d.o.f. --- as the dynamics of some nonlinear dispersive (quasi-) integrable background with respect to which a \textit{fast} d.o.f, $\underline{\phi}$, propagates linearly \cite{ref19,ref12}.
For concreteness and simplicity, notice that in the Cauchy description, we already obtain a Lax-pair formulation with a two-level dynamics: the 'slow' potential $V$ evolves along the KdV equation, on top of which the 'fast' master function $\phi$ propagates with respect to a wave-like equation 
\begin{equation}
\begin{cases}
    \big(-\partial_t^2+\partial_x^2-V\big)\phi=0 \hspace{0.1cm} (\text{in general, with any source term)}\\
    V_{,\sigma}-6VV_{,x}+V_{,xxx}=0
\end{cases} \hspace{0.1cm} .
\label{15}
\end{equation}
This is akin to the \textit{wave-mean flow} approach recently proposed in \cite{ref19,ref20,ref12} to account for the \textit{universality} and \textit{simplicity} of the binary BH merger waveform through underlying integrable structures.

\newpage 

\section*{Acknowledgments}

We warmly thank our advisors José-Luis Jaramillo, Carlos F. Sopuerta and Michele Lenzi, together with Jérémy Besson, Dimitrios Makris and Michel Semenov-Tian-Shansky for enlightening and fruitful discussions.


\end{document}